\documentclass[aps,prb,reprint,amsmath,amssymb,superscriptaddress,showpacs,floatfix]{revtex4-1}
\usepackage{graphicx,amsfonts,times,bm,amsmath,verbatim,color,array}
\usepackage{ifthen,braket,xcolor,bm}
\usepackage[colorlinks, allcolors=blue]{hyperref}
\usepackage{natbib}
\usepackage{braket}
\usepackage{cancel}                  

\begin{document}


\title{Fulde$-$Ferrel$-$Larkin$-$Ovchinnikov phase in one dimensional Fermi gas with attractive interactions and transverse spin-orbit coupling}


\author{Monalisa Singh Roy}
\email{singhrm@biu.ac.il, singhroy.monalisa@gmail.com}
\affiliation{Department of Physics, Bar-Ilan University, Ramat Gan 52900, Israel}
\author{Manoranjan Kumar}
\email{manoranjan.kumar@bose.res.in, manoranjan15@gmail.com}
\affiliation{S. N. Bose National Centre for Basic Sciences, Block - JD, Sector - III, Salt Lake, Kolkata - 700106, India}



\date{August 6, 2021}

\begin{abstract}
We examine the existence and characteristics of the exotic Fulde-Ferrel-Larkin-Ovchinnikov (FFLO) 
phase in a one dimensional Fermi gas with attractive Hubbard interactions, in the 
presence of spin-orbit coupling (SOC) and Zeeman field. We show that a robust FFLO phase can be created 
in the presence of attractive on-site interactions and Zeeman field, and that the addition of SOC 
suppresses the FFLO order and enhances the pair formation. In absence of SOC, the system shows four phases: 
Bardeen-Cooper-Schrieffer (BCS), FFLO, multi-mode pairing and fully polarized 
phases by tuning the Zeeman field $h$, and the quantum transition between these phases is 
discontinuous with respect to $h$. In the presence of SOC, the transition from the BCS to FFLO phase 
becomes continuous. We present a complete phase diagram of this model both in the presence and in the 
absence of SOC at quarter electron filling and also explore the effect of SOC on the FFLO phase.  
\end{abstract}

\maketitle

\section{Introduction} \label{Introduction}
The presence of external magnetic and electric fields in superconducting 
	materials give rise to many exotic phases and their effects have been extensively studied~\cite{2012_Wilson,2009_Abetti} since the discovery of superconductivity in 1911~\cite{1911_Onnes}. Over the years, discovery 
	of new superconducting materials and improvements in their synthesis mechanism have yielded steadily 
	increasing superconducting transition temperatures ($T_c$)~\cite{2011_Bhattacharya}, and more refined 
	applications of superconducting materials in daily life~\cite{book_Mele}. At high temperature, strong 
        magnetic field $h$ destroys the superconducting properties in materials~\cite{book_Annett}. Whereas, at low 
	temperatures and low to moderate magnetic fields, these materials give rise to many exotic phenomena like Meissner 
	effect~\cite{1933_Meissner} and Fulde-Ferrel~\cite{1964_Fulde} (FF) and 
	Larkin-Ovchinnikov~\cite{1964_Larkin,1965_Larkin} (LO) phases etc.
	In Bardeen-Cooper-Schrieffer (BCS) superconductors~\cite{1957_Bardeen}, electrons of opposite spins 
	and momenta form Cooper pairs, but in presence of low $h$ the Fermi energies of (\emph{up} spin and \emph{down} spin) electrons shift and 
	the electron pairing process gets affected. Fulde and Ferrel~\cite{1964_Fulde}, and Larkin and 
	Ovchinnikov~\cite{1964_Larkin,1965_Larkin} independently showed that in presence of magnetic field, 
	a robust superconducting order could co-exist with a magnetic order in superconductors, and electron 
        pairs with non-zero momentum can be formed
        in an inhomogeneous superfluid phase~\cite{2018_Kinnunen}. Since then, 
	there have been much efforts to realize this phase in various materials, especially in layered superconductors 
	like $\mathrm{L}\mathrm{a}_{2-x}\mathrm{B}\mathrm{a}_{x}\mathrm{C}\mathrm{u}\mathrm{O}_{4}$~\cite{2004_Fujita}, ${\mathrm{C}\mathrm{e}\mathrm{C}\mathrm{o}\mathrm{I}\mathrm{n}}_{5}$~\cite{2003_Bianchi}, and organic salts 
	like $\mathrm{B}\mathrm{E}\mathrm{D}\mathrm{T}-\mathrm{T}\mathrm{T}\mathrm{F}$~\cite{2007_Lortz}. However, this phase is fragile since any impurity or other perturbations 
	can disturb this phase in materials~\cite{2001_Agterberg,2007_Wang,Zuo_2009}.

In recent years, cold atoms confined in optical lattices have emerged as an excellent alternative playground 
to explore superconductivity in pristine conditions\--- to study different pairing mechanisms in it, and effects 
of various external fields on the superconducting state~\cite{2008_Bloch}. The existence of 
Bose-Einstein-Condensation (BEC) was demonstrated in a gas of cooled Sodium ($\mathrm{N}\mathrm{a}$) atoms by Davis {\it et al.}
in 1995~\cite{1995_Davis}. Since then, existence of superfluidity has been realized in various Fermi and 
Bose gases~\cite{2002_Greiner,2005_Lecheminant,2006_Chin,2009_Clade,2012_Desbuquois}. The physics of these 
gases trapped in optical lattices are well described by Hubbard like models with effective on-site 
interactions $U<0$ that are created by tuning Feshbach resonance in the system~\cite{2010_Chin}.
Synthetic spin-orbit coupling (SOC) and Zeeman fields are created through Raman coupling~\cite{2013_Galitski,2015_Zhai}. 
One dimensional (1D) Fermi gas with attractive interactions shows a BEC phase at very strongly attractive interactions, 
and a BCS phase with s-wave like pairings for moderate $U$~\cite{2010_Esslinger}. Introduction of Zeeman field 
$h$ in this system takes the system from a BCS phase at a low $h$, to a partially polarized phase at moderate $h$, 
and a fully polarized phase for high $h$~\cite{2010_Liao,2012_Feiguin}. This partially polarized phase 
at moderate $h$ is proposed to host exotic Fulde-Ferrel-Larkin-Ovchinnikov (FFLO) phase 
pairings~\cite{1964_Fulde,1964_Larkin,1965_Larkin}. This phase is characterized by finite momentum of 
the centre of mass of bound pairs, which manifests itself into twin peaks (at $\pm k$) 
in the pair density correlations in momentum space. 
	
The FFLO phase is more stable in 1D systems due to absence of eddy currents and phase separations 
which are more common in three dimensional (3D) systems and make it difficult for 3D systems 
to host the FFLO phase~\cite{2018_Kinnunen}. In addition, the 1D FFLO phase is expected to host the 
non-trivial p-wave like pairings in the presence of transverse SOC field~\cite{2014_Qu,2018_Ptok}. 
This phase is also proposed to host topological edge modes whose hallmarks are reported to be 
exponentially decaying energy gaps as a function of increasing system size~\cite{2015_Ruhman,2017_Ruhman}. There are  
many studies of model Hamiltonians, ranging from simple Fermi Hubbard models with additional 
interaction terms to systems with proximity induced superconducting terms, for exploring the existence of the FFLO phase. 
L{\"u}scher {\it et al.} studied 1D attractive Hubbard model in the presence of finite
spin polarization and showed the existence of FFLO phase and its fingerprint in 
spatial noise correlations~\cite{2008_Luscher}. Yang used a field theoretic approach to study 
the non-uniform superconducting states in quasi-1D systems and plotted a schematic 
phase diagram in the phase space of Luttinger liquid parameter $K$ and magnetic field $h$ \cite{2001_Yang}. Rizzi {\it et al.} also 
studied the attractive Hubbard model to study the stability of the FFLO phase in optical lattices~\cite{2008_Rizzi}. Feiguin 
{\it et al.} have studied this model with confining parabolic potential in the optical lattice~\cite{2012_Feiguin}. 
In this paper, we have presented systematic theoretical studies of quantum phase diagram of the 1D attractive Fermi gas model Hamiltonian subjected to Zeeman field and an SOC field, as a function of on-site interactions $U$.

In this paper, we study a simple model of 1D Fermi gas in the limit of attractive on-site 
interaction ($U<0$) to explore the FFLO phase and associated phase transitions at low filling fraction 
$\nu=0.25$. We study this system, both (i) in the absence of SOC, and (ii) in the presence of a transverse SOC. 
We find that the FFLO phase spans a large area of the phase diagram for all electronic densities, both 
in the absence and in the presence of SOC. We present a complete phase diagram of this model in the phase space 
of $U$ and $h$, and for SOC strengths $\alpha = 0$ and $0.05$. 

The paper is organized into four sections. In Sec.~\ref{Model and Method}, we introduce the model 
and the numerical technique. In Sec.~\ref{Results}, we discuss the main criteria used for identifying 
the FFLO phase in the system. We first focus on the case with no SOC, and discuss the different 
phases in the $h-U$ parameter space of the system. Next, we add a transverse SOC field (in the $x-$direction) and 
explore its effect on the FFLO phase. We conclude with a brief discussion of 
the reported results and their possible impact on the current understanding of exotic pairings in 
1D ultracold systems and implications thereof in Sec.~\ref{Discussion}.

\section{Model and Method} \label{Model and Method}

We study the 1D Fermi gas with attractive on-site interactions $U$, in presence of a Zeeman field $h$  
and transverse SOC field $\alpha$. The model Hamiltonian of this system can be written as,
\begin{equation} \label{eq.1: Hamiltonian}
 H = H_{\text{\normalfont t}} + H_{\text{\normalfont U}} + H_{\text{\normalfont Z}} + H_{\text{\normalfont SOC}} ,
\end{equation}
where,
\begin{align*}
 H_{\text{\normalfont t}} = & -t \sum_{i, \sigma} \left( C_{i, \sigma}^{\dagger} C_{{i+1, \sigma}} + h.c. \right)\\ \nonumber
H_{\text{\normalfont U}} = & U \sum_i  n_{i, \uparrow} n_{i, \downarrow} \text{\normalfont,} \hspace{3ex} 
H_{\text{\normalfont Z}} = -h \sum_{i} S_{i}^{z}\text{\normalfont,} \\ \nonumber
 H_{\text{\normalfont SOC}} = & + i \alpha \sum_{i} \left( C_{i, \uparrow}^{\dagger} C_{i+1, \downarrow}  + C_{i, \downarrow}^{\dagger} C_{i+1, \uparrow} - h.c. \right) \text{\normalfont ,} \\ \nonumber
\end{align*}
where, $C_{i,\sigma}^{\dagger}$ $(C_{i,\sigma})$ are creation (annihilation) operators and $n_{i,\sigma}$ is the number 
operator at site $i$, with $\sigma=\uparrow$ up spin or $\downarrow$ down spin. $t=1$ defines the energy scale for our calculations. We 
 study the system away from the half-filling limit and in the attractive interaction regime $U \in [-1,-4]$. 
The quantity $\nu=n/2N$ defines the filling fraction of a system of $N$ sites containing $n$ electrons. In the absence of $U$ 
and $\alpha$, the spin up and spin down electronic bands split in the presence of the external magnetic field $h$. 
An attractive $U$ induces intra-band pairing correlations, whereas a transverse SOC generates spin momentum locking 
along the $x$-axis of the system, thus giving rise to a p-wave like pairing~\cite{kitaev2001unpaired}.

We have used the DMRG method for solving the Hamiltonian in Eq.~\eqref{eq.1: Hamiltonian}.
It is a state-of-the-art numerical technique for accurately calculating the low-lying eigenvalues and 
eigenvectors of low dimensional systems~\cite{1998_White,2005_Schollwock}. 
It is based on the systematic truncation of irrelevant degrees of freedom at every step of the infinite 
and finite DMRG algorithms. In the fermionic system under study, the spin degrees of freedom are not conserved, 
hence the Hamiltonian dimension is significantly large. The eigenvectors of the density matrix of the 
system block corresponding to $m \simeq 700$ largest eigenvalues have been retained to maintain a reliable 
accuracy. More than $10$ finite DMRG sweeps have been performed for each calculation to minimize the error 
in per site energies to less than $1\%$ and the maximum system size studied is $N=120$.


\section{Results} \label{Results}

In this section we present the numerical studies investigating the existence of a 
robust FFLO phase in a 1D ultracold atoms with intrinsic attractive on-site 
interactions and Zeeman field, first in the absence of SOC and then in presence of a 
transverse SOC field in $x-$direction. 
Most of the results presented in this paper is 
for the quarter filling fraction $\nu=0.25$, however, we found 
that the FFLO phase exists for a wide range of densities. We also present 
a quantum phase diagram of the model in $h-U$ parameter space for $\nu=0.25$ and study the effect of 
finite SOC interactions $\alpha >0$ on the FFLO phase. In this paper, we show that 
a robust FFLO phase exists for a wide range of electron fillings $U$, $h$ and $\alpha$, 
and this phase is distinct from other phases like the BCS and a multi-mode pairing 
(MMP) phase, which appears just before the system transitions into a fully polarized 
(FP) phase at high $h$. 

To characterize various phases in the system we study the pair density correlations or singlet pair 
correlation function $P(i,j)$ and its Fourier transform (F.T.). The singlet pair correlation is defined as, 
\begin{equation}
P(i,j) = \langle C_{i \uparrow}^{\dagger} C_{i\downarrow}^{\dagger} C_{j \uparrow} C_{j \downarrow} \rangle \text{\normalfont.} 
\end{equation}
Its F.T. of $P(i,j)$ is given by,
\begin{equation}
P(k) = \sum_{k} e^{- i {\bf k} \cdot {\bf r}_{ij}} P(i,j)\text{,} 
\end{equation}
where $(r_{ij} = i - j )$, and $i, j$ represent site indices in the system.
The single peak in $P(k)$ at $k=0$ indicate that the electron pairs are formed at zero momentum, 
i.e., Cooper pair formation at $k=0$, which is a signature of the BCS phase. Twin peaks at finite 
momenta in $P(k)$ momentum distribution curve is a hallmark of an underlying FFLO-like pairing 
where electron pairs are formed with a finite momentum. In a mixed BCS-FFLO 
phase, where both the conventional BCS phase and the FFLO phase co-exist, $P(k)$ shows three peaks \--- one at 
zero momentum $k=0$, and twin peaks at a finite momentum $k_{h}$. At sufficiently high magnetic field 
$P(i,j)$ is short range in nature, and $P(k)$ shows two peaks at $\pm k_{h}$ and a constant value of $P(k)$ between 
these two momenta, i.e., all the momenta between $\pm k_{h}$ contribute equally. We call this phase as MMP phase.  
In a fully polarised phase, $P(k)$ is zero. We also study two energy gaps \--- the pair binding 
energy ($E_b$) or the parity gap, and the excitation energy gap $\Delta$, defined as,
\begin{subequations}
\begin{align}
E_b (n,N) & = \dfrac{1}{2} \left[ E_{0}(n+1,N) + E_{0}(n-1,N) - 2 E_{0}(n,N) \right]\\
\Delta (n,N) &=E_{1} (n,N) - E_{0} (n,N) \text{,}  \label{eq.2: Excitation Gap}
\end{align}
\end{subequations}
where, $E_{0}(n,N)$ and $E_{1}(n,N)$ represent the ground state energy and the 
first excited state energy, respectively, with $n$ electrons in the system of $N$ sites. The finite binding energy $E_b$ 
is the signature of the BCS phase, whereas the exponential decay of $\Delta (n,N)$ may indicate 
the existence of a topological phase.      

\begin{figure}
\centering
\includegraphics[width=0.98\linewidth]{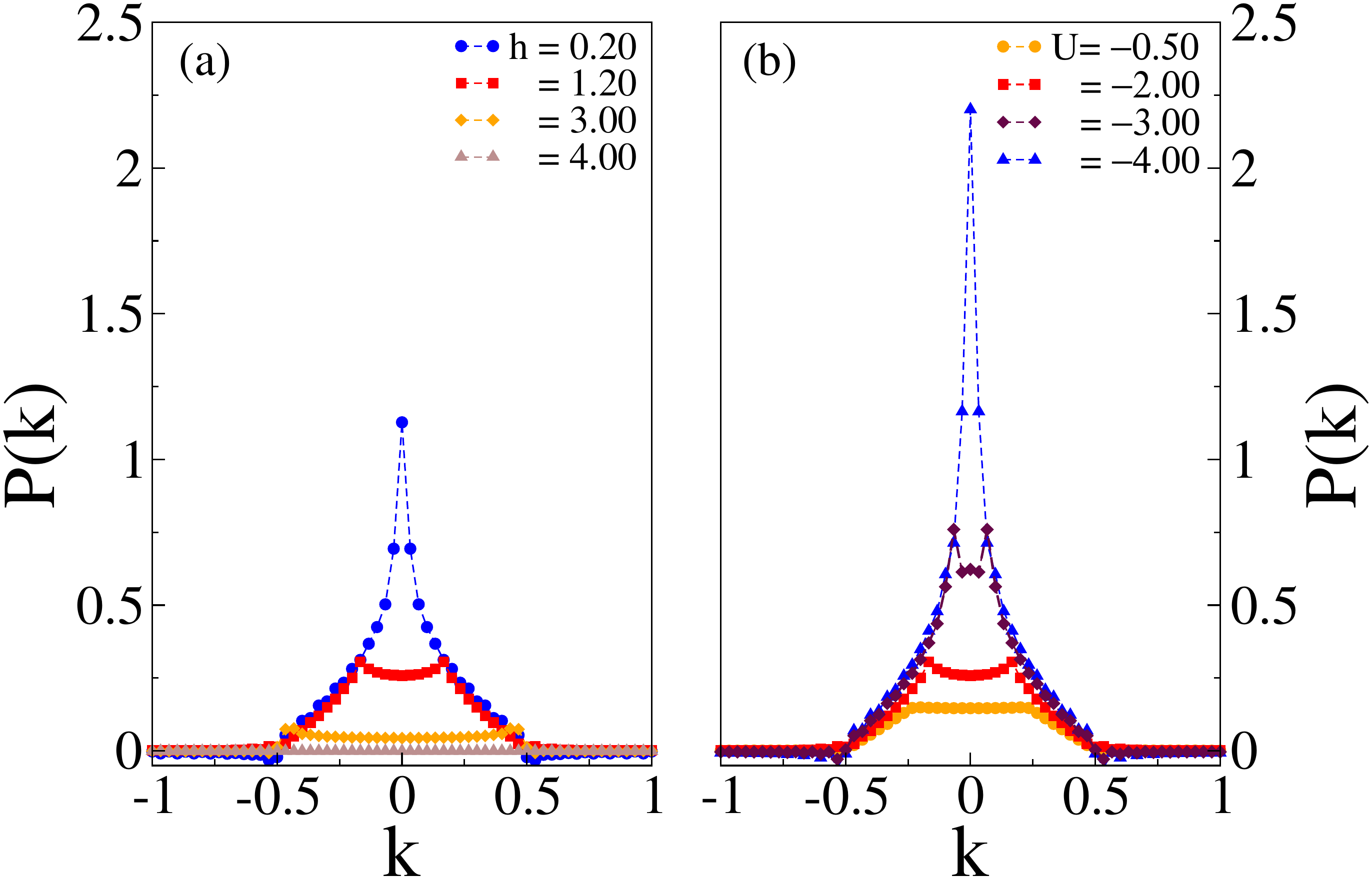}
\caption{\small F.T. of singlet pair density colrrelations, $P(k)$ vs. $k$  for
(a) different $h=0.20$, $1.20$, $3.00$, and $4.00$ at $U=-2.00$, and for
(b) different $U=-0.50,-2.00,-3.00$ and $-4.00$ at $h=1.20$,
for $\nu=0.25$, $\alpha=0$.}  \label{fig.1}
\end{figure}

The other calculated quantities include local charge density with up spin $n_{\uparrow}(i)$ and down spin $n_{\downarrow}(i)$, 
and local spin density $S^z(i)$. We notice that the spatial distributions of these quantities are different for 
each of the observed phases. In the BCS phase $n_{\uparrow}(i)$ and $n_{\downarrow}(i)$ have overlapping spiral nature.
For the partially polarized phases \--- FFLO and MMP, $n_{\uparrow}(i)$ and $n_{\downarrow}(i)$ are separated and difference in 
the wavelength or pitch angle of the spiral oscillations are observed. 


\subsection{In Absence of SOC} \label{SUB.alp0}

In absence of any magnetic field, the system remains in the trivial BCS phase which is characterized 
by Cooper pairs with zero net momentum. In presence of finite magnetic field $h$, the system can transition 
into the FFLO phase, where the electron pairs are formed with net non-zero momentum in presence of finite magnetic field $h$. 
This phase is expected to retain quasi-long range correlations, especially in 1D. 
To ascertain that, in Fig.~\ref{fig.1} the F.T. of singlet-pair density correlation, $P(k)$ vs. $k$, are 
plotted for different $h$ and $U$, for $\nu=0.25$ and in absence of SOC, i.e., $\alpha=0$. It shows 
a single peaked structure at $k=0$ at low $h$, as expected from a trivial BCS phase. Increasing $h$, 
two-peaked $P(k)$ with maxima at $\pm k_{h}$ are observed, indicating the presence of an FFLO phase. 
In the fully polarized (FP) phase at high $h$, $P(k)$ is vanishingly small and no peak is observed 
in $P(k)$. Between the FFLO and FP phase, there exists a narrow regime of the MMP phase. In this phase, 
$P(k)$ shows a plateau like structure between the twin peaks, i.e., the $P(k)$ is uniformly distributed 
between the momenta $\pm k_{h}$, which indicates that the momentum of centre of mass of the condensate is 
distributed between $ \pm k_{h}$ around the Fermi-momentum. The $P(i,j)$ is fast and algebraically 
decaying function. The peak height of $P(k)$ is significantly smaller in this phase, as compared to the FFLO phase. 

\begin{figure}
\centering
\includegraphics[width=0.9\linewidth]{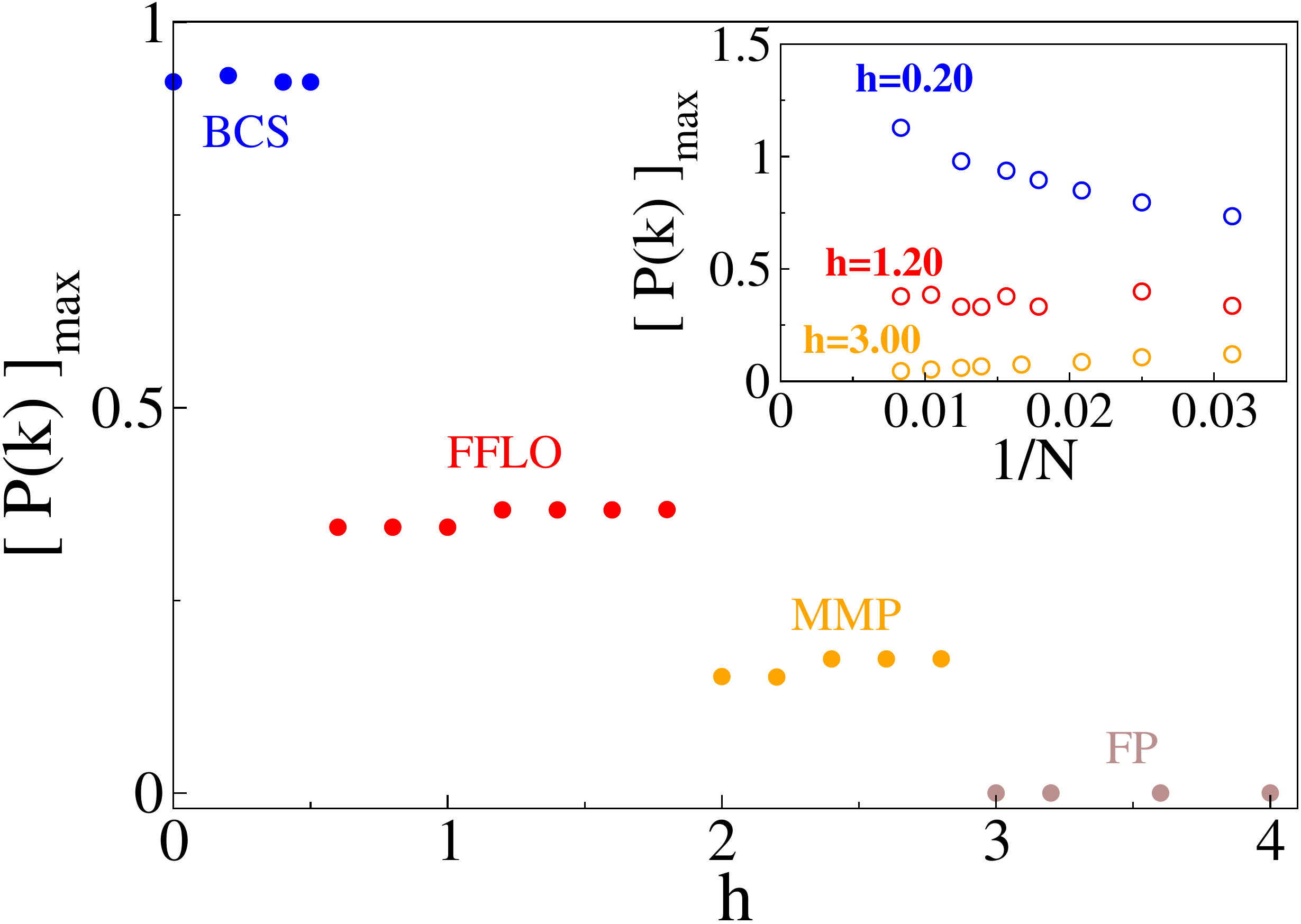}
\caption{\small The maxima of F.T. of singlet pair density correlations, $\left[P(k) \right]_{max}$ as a function of $h$, at $U=-2.00$, $\nu=0.25$, and $\alpha=0$. Inset: $\left[P(k) \right]_{max}$ as a function of inverse system size $1/N$ for different $h$, at $U=-2.00$, $\nu=0.25$, and $\alpha=0$.} \label{fig.2}
\end{figure}

In Fig.~\ref{fig.1} (a), variation of $P(k)$ for four values of $h$ are shown for $U=-2.00$. In Fig.~\ref{fig.1} 
(b) the magnetic field $h=1.20$ is kept fixed and $P(k)$ is plotted for four values of $U$. Larger value of $|U|$ 
increases binding energy, therefore, BCS phase is favoured at higher magnitude of $U$, and FFLO-like pairing is observed at 
weakly attractive $U$ at comparatively lower $h$. For larger $|U|$, larger magnetic field is required to break 
the bound electron pairs, hence we notice that $U=-0.50$ is already in the FP state for the given $h$. Whereas, 
for $U=-4.00$ the BCS phase remains intact upto a large $h$. We also plot the peak height of $\left[ P_{max}  (k) \right]$ 
as a function of $h$ for $U=-2.00$ in Fig.~\ref{fig.2}, and find that the four phases in this system can be easily identified by 
the respective plateaus in $\left[ P_{max} (k) \right]$ corresponding to each phase. We also notice that transition 
from one to the other phase occurs through discontinuous jumps. We plotted $\left[ P_{max}(k) \right]$ as a function 
of $1/N$ in the inset of Fig.~\ref{fig.2}, and notice that the effect of the finite size is weak in the FFLO and MMP phases, 
but $\left[ P_{max} (k) \right] $ increases with the system size in the BCS phase. 

\begin{figure}
\centering
\includegraphics[width=0.98\linewidth]{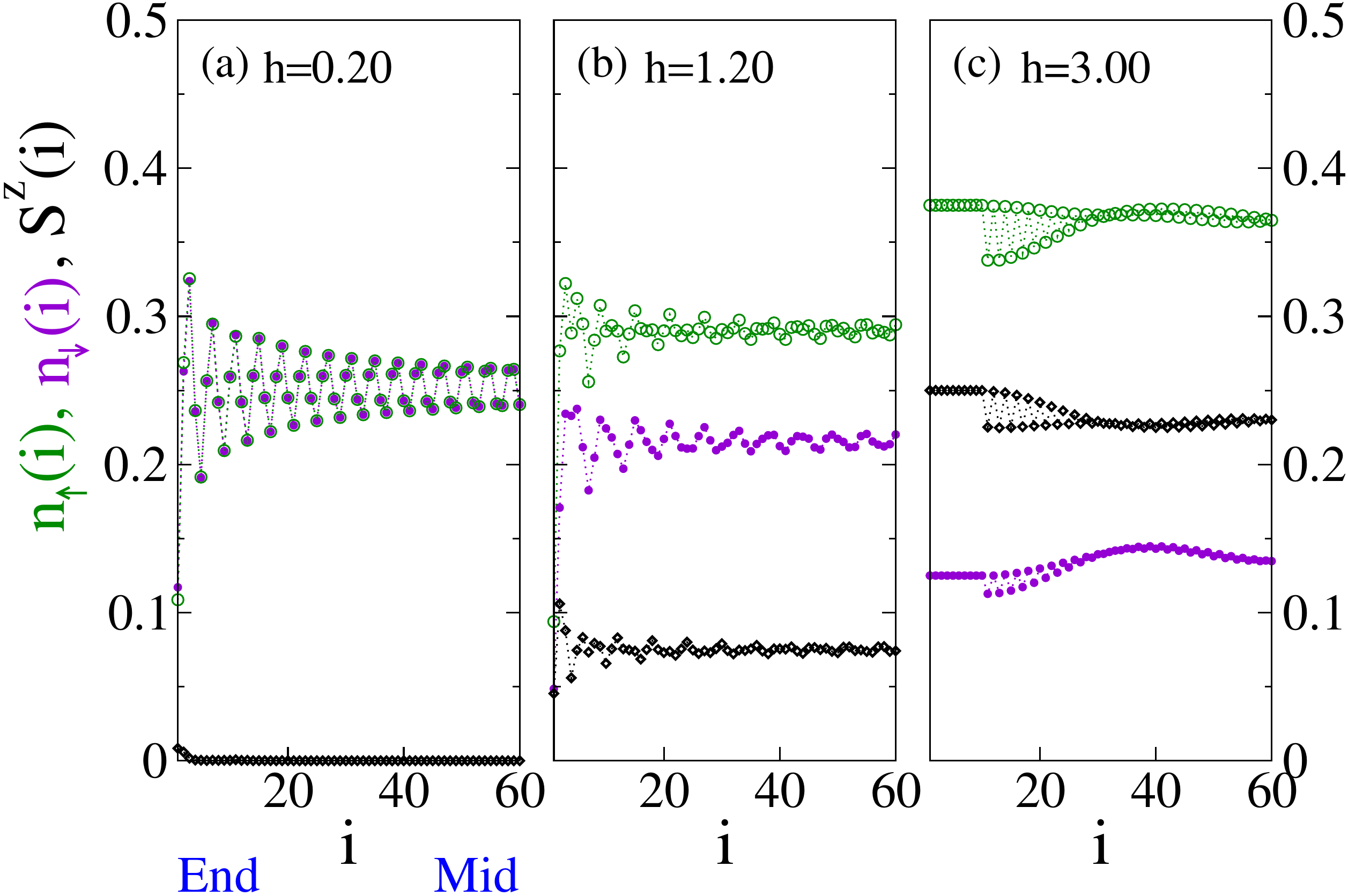} 
\caption{\small Spatial profile of local up charge density $n_{\uparrow}(i)$, local down charge density $n_{\downarrow}(i)$, and local spin density $S^z(i)$ at (a) $h=0.20$, (b) $h=1.20$, and (c) $h=3.00$, , for $U=-2.00$, $\nu=0.25$, $\alpha=0$ and $N=120$.} \label{fig.3}
\end{figure}

For further understanding of different phases, the behavior of local charge and spin densities 
are analyzed for three values of magnetic field $h$, 
corresponding to the three phases 
at $U=-2.00$. We ignore the fully polarised regime where all spins are polarized along the 
direction of magnetic field $h$ and the charge is uniformly distributed. 
Fig.~\ref{fig.3} shows the spatial profile of the spin densities $n_{\sigma}(i)$ and local magnetization, 
$S^z(i)=n_{\uparrow}(i) - n_{\downarrow}(i)$, for different $h$. At low $h=0.20$ (Fig.~\ref{fig.3}(a) ), 
the up and down spin densities overlap and the system is in a trivial BCS phase, which is a non-magnetic state.
Above a threshold magnetic field $h_{c_1}$ some of the singlet pairs are broken, leading to 
a partial magnetic polarization $S^z(i)\ne 0$ in the system as shown in Figs.~\ref{fig.2} (b) and (c). 
The charge density wave (CDW) oscillations have a maximum amplitude for low $h=1.20$ (Fig.~\ref{fig.3} (a)) 
and its amplitude decreases with increasing $h$ (Figs.~\ref{fig.3} (b) and (c)). 
At $h=3.00$ (Figs.~\ref{fig.3} (c)), the system is in the MMP phase and the density modulations vanish at the mid 
of the chain. Above another threshold value of magnetic field $h_{c_2}$, the system transitions from the MMP phase to the FP phase. 

Further analysing the oscillations in the local charge density, we find that at low $h$, where 
the system is in a BCS phase, oscillation in $n_{\sigma}(i)$ are described 
by a sinusoidal function with its amplitude decaying from the edges towards the centre. The functional 
form of $n(i)_{\sigma}$ in this regime is given by: $A sin(\gamma x + A_0) x^{-\eta} + C$. The charge density 
profile does not change appreciably with increasing $h$ in the BCS phase. For $h>h_{c_1}$, another 
sinusoidal length scale sets in for the partially polarized phases \--- FFLO and MMP phases, 
and $n(i)$ can be fitted with the 
charge density profile: $A sin(\gamma x + A_0) sin(\beta x+B_0) x^{-\eta} + C$. 
The power law exponent remains $\eta \sim 1$ for all $h<h_{c_2}$. Whereas the wavelength of one 
of the sinusoidal function $\lambda_2 = \dfrac{2 \pi}{\beta}$ decreases with increasing $h$,
the wavelength of the other sinusoidal $\lambda_1 = \dfrac{2 \pi}{\gamma}$ does not vary with $h$ in the FFLO  phase.
In the MMP phase the wavelength $\lambda_2 = \dfrac{2 \pi}{\beta}$ becomes very large, whereas the 
other wavelength $\lambda_1 = \dfrac{2 \pi}{\gamma}$ decreases significantly to approximately $\sim 2-4$ lattice units. 
\begin{figure}
\includegraphics[width=0.8\linewidth]{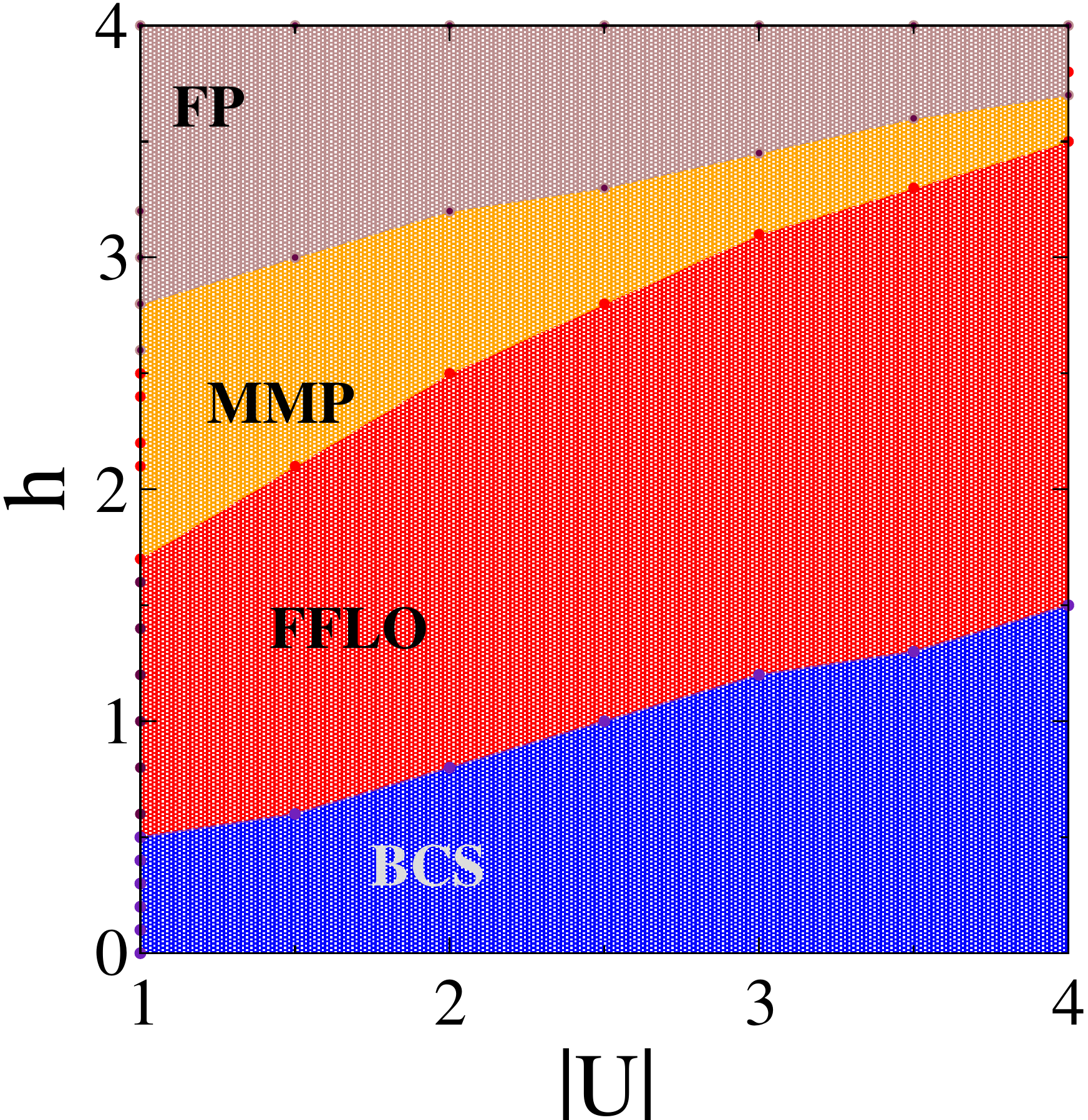} 
\caption{\small Phase diagram of the model described by Eq.~\eqref{eq.1: Hamiltonian}, in the phase space of magnetic field $h$ and on-site attractive interactions, $\vert U \vert$, at $\alpha=0$ and $\nu=0.25$.} \label{fig.4}
\end{figure}

In Fig.~\ref{fig.4}, we present the quantum phase diagram of the model Hamiltonian described 
by Eq.~\eqref{eq.1: Hamiltonian}, in absence of SOC ($\alpha=0$) and for $\nu=0.25$, based 
on information extracted from $ P(k)$, their maxima, charge and spin density profiles. All the 
phase boundaries are based on the $N=96$ system size and we note that the finite size scaling of the 
boundary is very weak. 
In absence of any $h$, BCS phase is observed for all values of $|U|$. 
Upon increasing $h$, the system goes from a BCS phase to the partially polarized FFLO phase, 
then from the FFLO phase to the MMP phase, and finally to the FP phase at high $h$. The FFLO phase is a dominant 
phase in the quantum phase diagram and the width of this phase increases with $|U|$. The magnetic 
field required for the transition from the BCS to the FFLO phase, $h_{c_1}$, is smooth and linear with $U$. Similarly, 
the $h$ required for the FFLO to the MMP phase transition also varies linearly with the $\vert U \vert$, especially 
in small $U <2.5$ limit. We also explored the quantum phase diagram at lower fillings $\nu$ 
(the lowest filling studied was $\nu=0.10$), and found that
the phase boundaries shift towards lower $h$, i.e., slope of the $h_c$ curves increases with decreasing 
electron filling $\nu <0.25$. This is because, at lower densities, lower $h$ is sufficient to break 
the bound pairs in the system.


\subsection{In Presence of SOC} \label{SUB.}

In this section we explore the effect of small SOC strength $\alpha=0.05$, which is expected 
to produce a p-wave pairing in superconducting phase for attractive $U$ interactions. This 
exotic p-wave like phase is proposed to host topological edge modes \cite{2018_Ptok,2020_Singh_Roy}, which 
is important for applications in quantum computation. In the superconducting BCS state, the binding energy of 
the system should be finite, whereas in the FFLO, the MMP and the FP phase unpaired electron 
should have zero binding energy.

\begin{figure}
\centering
\includegraphics[width=0.9\linewidth]{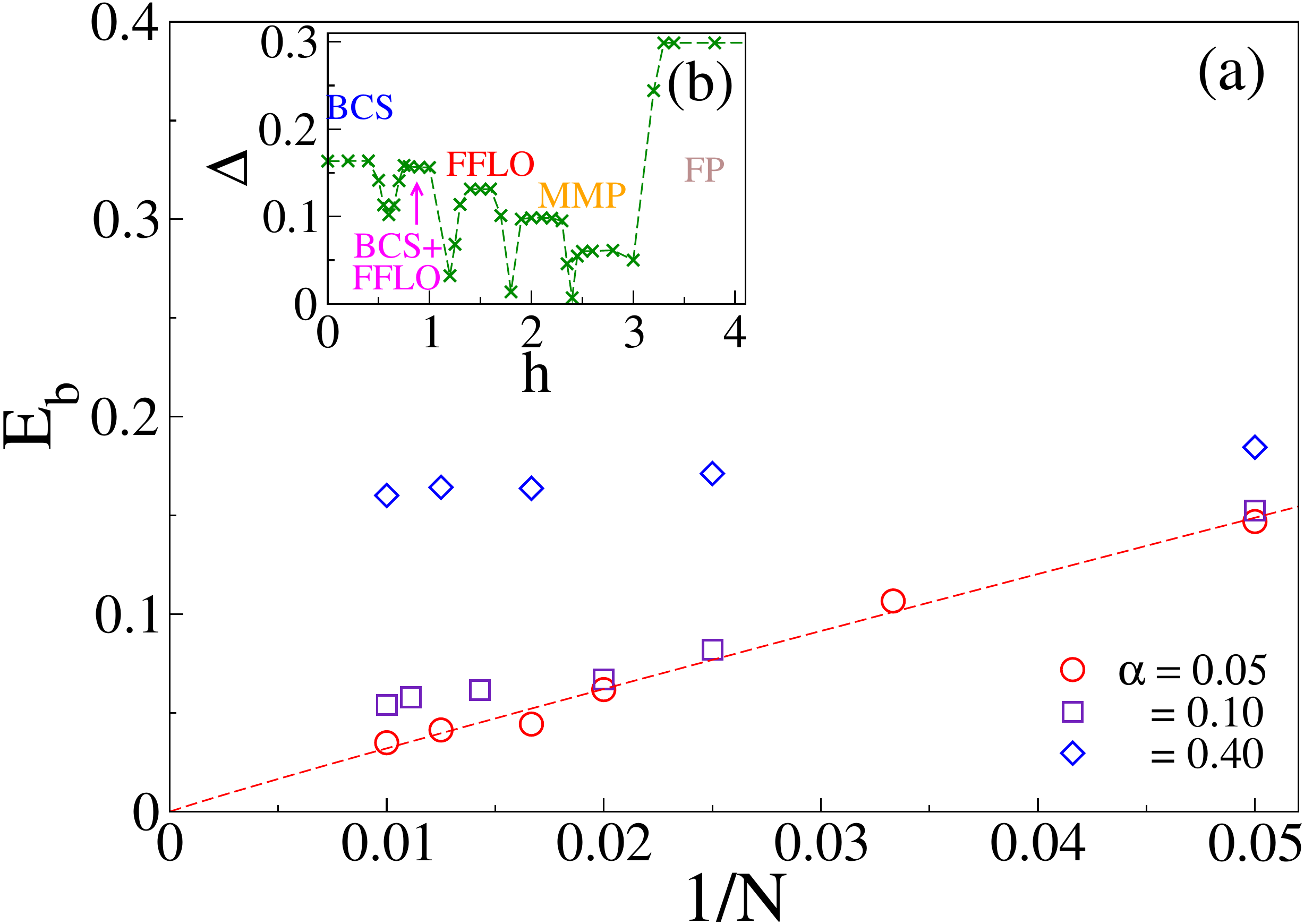}
\caption{\small Binding energy as function of $N$ for different $\alpha$, at $U=-1.00$, $\nu=0.25$ and $h=1.00$. Inset: First excitation energy gap $\Delta$  as a function of $h$ for $\alpha=0.05$, at $U=-2.00$, and $\nu=0.25$} \label{fig.5}
\end{figure}

\begin{figure}[t]
\centering
\includegraphics[width=0.9\linewidth]{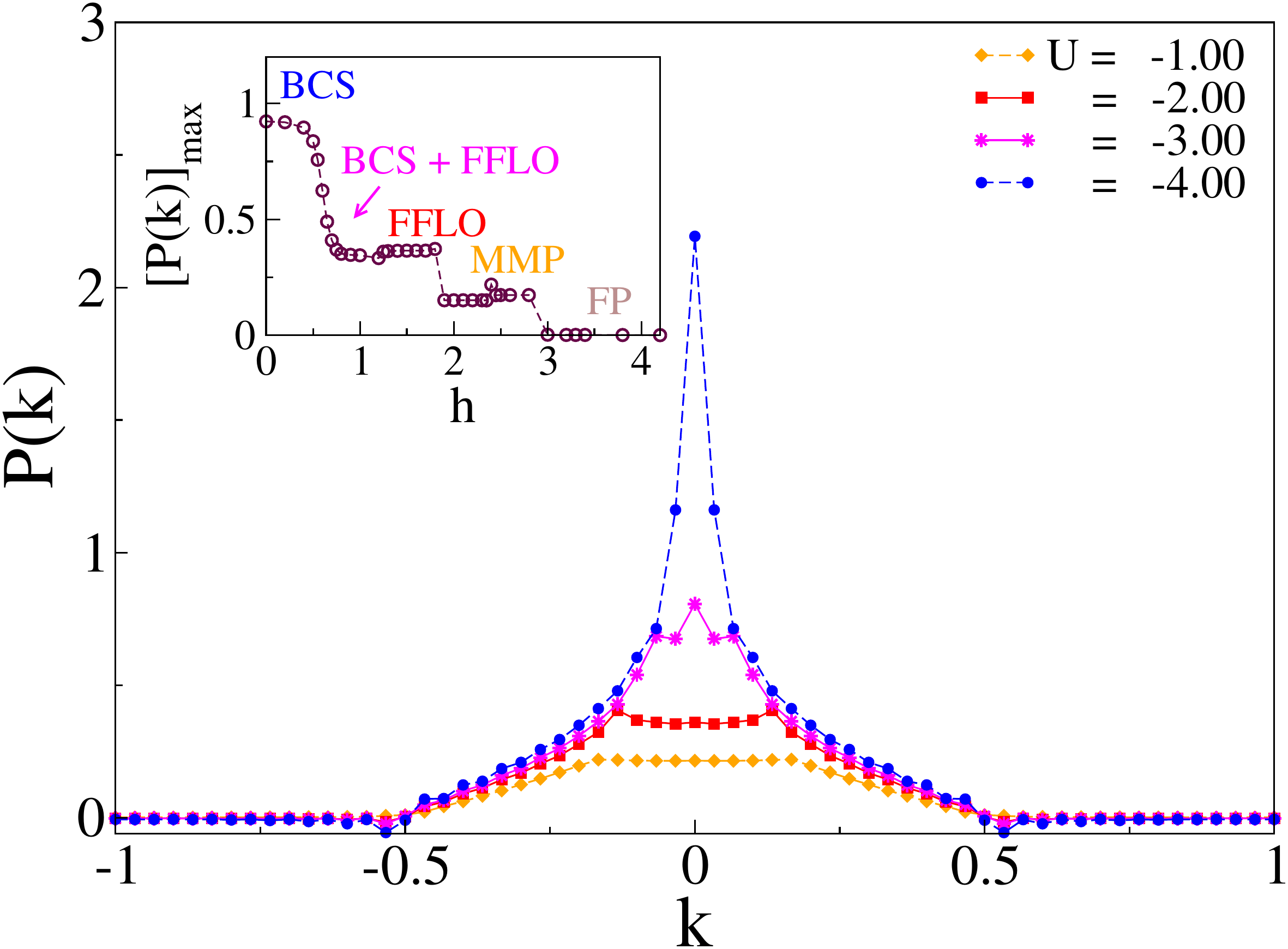}
\caption{\small F.T. of singlet pair density colrrelations, $P(k)$ vs. $k$ for different $U$, at $h=1.00$, $\nu=0.25$, $\alpha=0.05$. Inset: The maxima of F.T. of singlet pair density correlations, $\left[P(k) \right]_{max}$ as a function of $h$, at $U=-2.00$, $\nu=0.25$, and $\alpha=0.05$.} \label{fig.6}
\end{figure}

The binding energy is plotted as a function of system size for different $\alpha$ in  
Fig.~\ref{fig.5} for $\alpha=0, 0.05, 0.10,$ and $0.40$ at $\nu=0.25$ for $h=1$. We find that $E_b$ vanishes 
algebraically with $N$ for low $\alpha$, which corresponds to the FFLO phase. For higher $\alpha$, system goes 
to the BCS phase and $E_b$ has finite value in the thermodynamic limit.
The inset of Fig.~\ref{fig.5} shows the variation of the lowest excitation energy 
gap $\Delta$ with $h$ . The inset of Fig.~\ref{fig.5} shows fluctuations at the phase boundaries of the $\Delta-h$ plot, for a given system size. The fluctuations at the boundary can be utilised to determine the phase boundaries 
and we notice that the boundaries determined by this method agree well with 
those indicated by the pair correlation structure factor $P(k)$ in Fig.~\ref{fig.6}.  

\begin{figure}[t]
\includegraphics[width=0.9\linewidth]{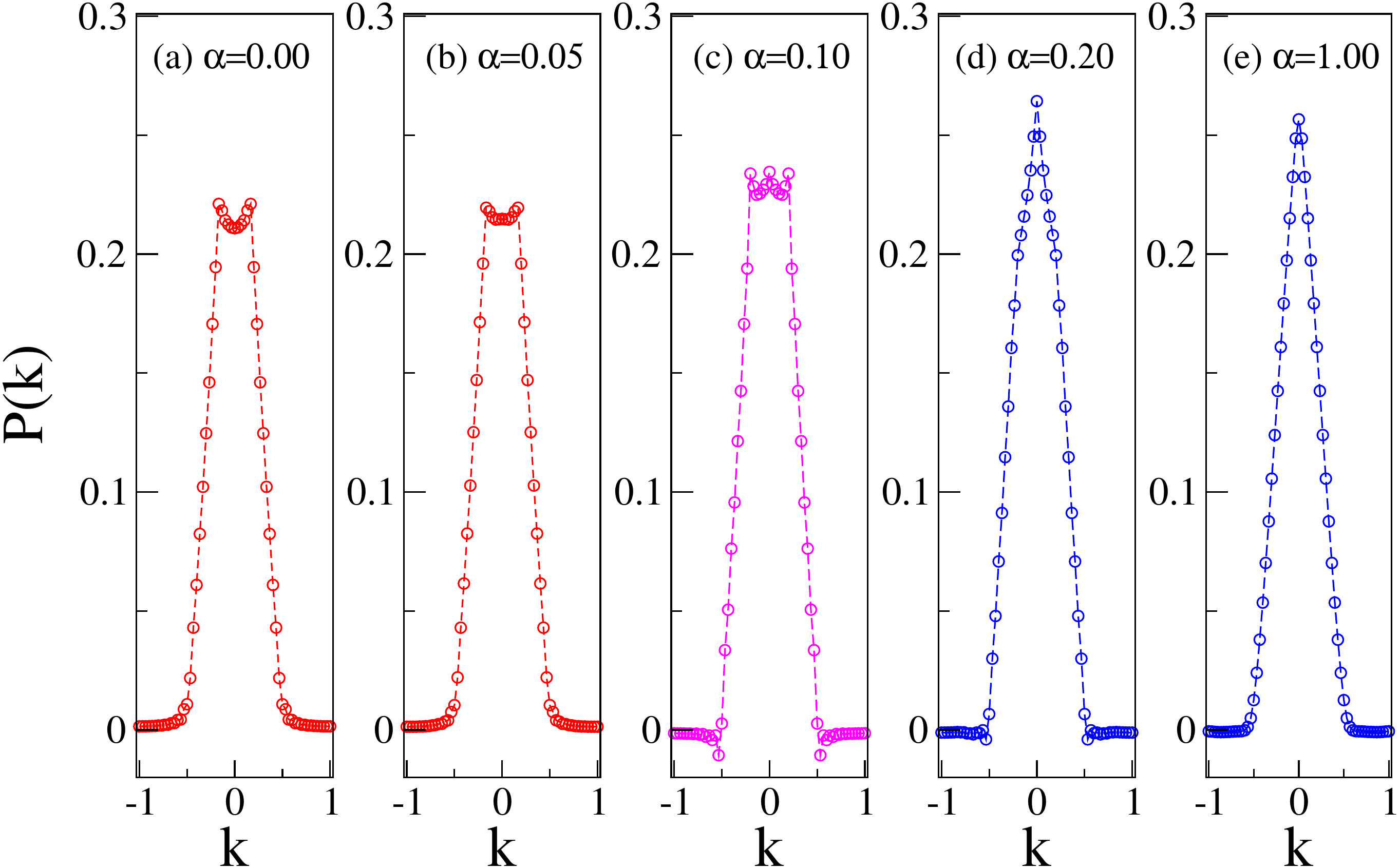}
\caption{\small F.T. of singlet pair density colrrelations, $P(k)$ vs. $k$ for different $\alpha$, at $h=1.00$, $\nu=0.25$, $U=-1.00$.} \label{fig.7}
\end{figure}

$P(k)$ vs. $k$ is now plotted in Fig.~\ref{fig.6}, for four values of $U$ at $h=1.00$, $\alpha=0.05$ and $\nu=0.25$. 
The FFLO pairings are observed for weakly attractive $U$, as indicated by the twin peaks in $P(k)$ 
for $U=-1.00$ and $-2.00$, whereas, a single peaked $P(k)$ is observed for stronger $U=-4.00$, indicative of 
a BCS phase. At $U=-3.00$, a three-peaked structure is observed, which is signature of an exotic mixture 
of BCS and FFLO phases. It should be noted that this exotic mixture state is not observed for any value 
of $h$ and $U$ in absence of $\alpha$. The $[P(k)]_{max}]$ is plotted as function of $h$ as shown in the inset of Fig.~\ref{fig.6}. 
Contrasting with the inset of Fig.~\ref{fig.2}, we find that the transition from BCS to FFLO phase 
now shows a smooth transition through a mixed BCS-FFLO phase, which was earlier discontinuous 
in absence of SOC. The transition from FFLO to the MMP phase and from MMP to FP 
phase remains discontinuous with increasing $h$, as before. We also checked the dependence of electronic 
filling for $\nu \in \left[0.10,0.25\right]$ and found that the FP phase is reached for lower $h$ at lower filling 
but the qualitative behaviour and sequence of the phases remains the same as the no SOC case. 

To understand the effect of $\alpha$ on the system, we study the $P(k)$ vs. $k$ 
characteristics for different strengths of $\alpha$ in Fig.~\ref{fig.7}. We find that the FFLO phase 
is retained at low $\alpha$ (Figs.~\ref{fig.7} (a) and (b) ), and the BCS phase sets in for 
higher $\alpha$ (Fig.~\ref{fig.7}(d) and (e) ), for a fixed $h$. For intermediate $\alpha$ (Fig.~\ref{fig.7}(c) ), a 
mixed FFLO-BCS phase is observed. 

{A quantum phase diagram of this system for a fixed $\alpha=0.05$ is shown in Fig.~\ref{fig.8} based on various criteria. 
It shows that in absence of $h$, BCS phase is observed for all attractive $U$. Upon increasing $h$, the 
system goes from first an unpolarized BCS phase to a partially polarized FFLO phase continuously, through 
a mixed BCS-FFLO phase \--- which was earlier not observed in absence of an SOC field (Fig.~\ref{fig.4}). The width 
of the BCS phase and FFLO phase shrinks in presence of the SOC. Thereafter, increasing $h$ the 
system goes from the FFLO to the MMP phase. 
The width of MMP phase increases in large $|U|$ limit, in the presence of SOC. Further enhancing $h$ lead to the FP phase, and width of this 
phase increases in smaller $|U|$ in presence of SOC. We checked that a similar phase diagram is observed 
for lower fillings $\nu$ as well, except that the phase boundaries are shifted towards lower $h$. For higher 
$\alpha$, the phase boundaries are shifted towards higher $h$.

\begin{figure}[t]
\includegraphics[width=0.8\linewidth]{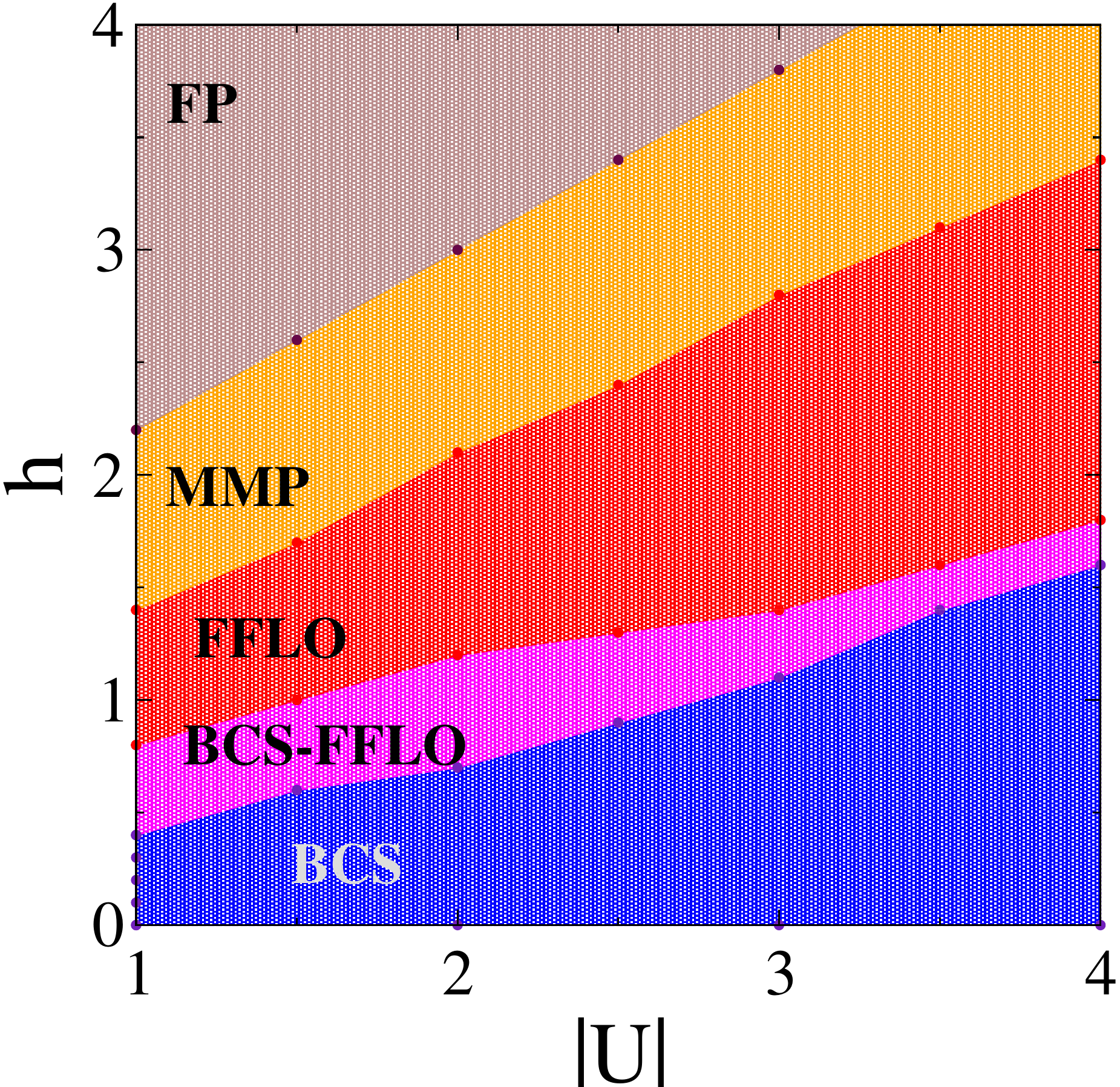}
\caption{\small Phase diagram of the model described by Eq.~\eqref{eq.1: Hamiltonian}, in the phase space of magnetic field $h$ and on-site attractive interactions, $\vert U \vert$, at $\alpha=0.05$ and $\nu=0.25$.} \label{fig.8}
\end{figure}


\section{Discussion and  Conclusion} \label{Discussion}

In this work we study the 1D Fermi gas model Hamiltonian described by Eq.~\eqref{eq.1: Hamiltonian} 
with attractive on-site interactions, SOC, and a Zeeman field, and explore the existence and signatures 
of the exotic FFLO phase in this system. We present the quantum phase diagram of this model in the $h - \vert U \vert$ parameter 
space, both in absence and in presence of SOC, at electron filling $\nu=0.25$. Most of the earlier 
works have explored the FFLO phase in 1D system at $\nu=0.50$~\cite{2007_Feiguin} and 
$\nu=0.25$~\cite{2015_Liang} and restricted their studies to just 
characterizing the FFLO phase~\cite{2018_Kinnunen}. 
We have found the quantum phase diagram of this model (Eq.~\eqref{eq.1: Hamiltonian}) in the phase space of $h - \vert U \vert$ for different $\alpha$, and provided different criteria to identify the phases observed in this system, including the different partially polarized phases at $\nu=0.25$. We note that the sequence of phases are the same for lower fillings as well, at least upto $\nu=0.10$.

We find that on-site interactions $U<0$ and SOC interactions $\alpha$ promote BCS pairings, whereas 
the Zeeman field $h$ promotes the FFLO order in the system. In the $h-\vert U \vert$ phase space 
we find four different phases: (i) the BCS phase, (ii) the FFLO phase, (iii) the MMP phase, 
and (iv) the FP phase. In the trivial BCS phase all the 
electrons form singlet (Cooper) pairs and have charge density oscillations at low $h$, 
low $\alpha$ and weak $U$. In the exotic FFLO and MMP phases, commensurate charge 
and spin oscillations at moderate $h$, moderate $U$, and moderate $\alpha$ are noted.  
The FFLO phase is characterized by quasi-long range order in the system 
and correlated singlet pairs with finite momenta~\cite{2001_Yang}, which is reflected through 
twin-peaks in the F.T. of singlet pair density correlations in the system. We find that 
both in absence and in presence of SOC field, the FFLO phase occupies a large area of the 
quantum phase diagram.
We also report a new quantum MMP phase which phase exhibits a short range singlet pair density $P(i,j)$ ordering. 
In this phase $P(k)$ shows a plateau like behaviour between two maxima, i.e, the centre 
of mass of the electron pairs can take any value between the $\pm k_{h}$. It is very different from 
the FFLO phase where the centre of mass of the electron pairs lies at $\pm k_{h}$, 
i.e., the momenta where the peaks are sharply defined. In presence of SOC, a mixed state exhibiting 
both BCS and FFLO pairings is also observed. The phase diagram of the system remains similar for lower 
electronic fillings $\nu \in \left[0.10,0.25\right]$, except that the phase boundaries shift to lower $h$ for lower $\nu$. 

In summary, this work presents a comprehensive study of the quantum phase diagram 
of a 1D Fermi gas system with instrinsic attractive on site interactions, a 
Zeeman field, and transverse SOC, and discusses in details various methods of characterizing 
these phases in similar 1D Fermi gas systems. We show that the FFLO phase dominates 
the phase diagram and it is robust even in presence of the SOC. This could have potential applications 
in understanding the unconventional superconductivity phases in low dimensional electron gas. 
This model can be easily implemented in the trapped cold atoms in optical lattices.

\begin{acknowledgements}\label{Acknowledgement}
MSR thanks Prof.~Fabian~Heidrich-Meisner for discussion. MK thanks SERB for financial support through grant sanction number CRG/2020/000754.
\end{acknowledgements}

\label{References}





%


\end{document}